\documentstyle[12pt]{article}
\begin{document}

\vspace*{-14mm}

\hfill hep-th/0007094

\hfill May 2001 (revised)

\thispagestyle{empty}

\vspace*{4mm}

\begin{center}

{\LARGE \bf  Matching the observational value of the cosmological
constant}\footnote{This paper is dedicated to the memory of
H.B.G. Casimir. \ Reported at the Marcel Grossmann Meeting, MG IX MM,
Rome, July 2000.}

\vspace{4mm}

\medskip

{\sc E. Elizalde}\footnote{Presently on leave at Department of Mathematics,
Massachusetts Institute of Technology, 77 Massachusetts Ave, Cambridge,
MA 02139. E-mail: elizalde@math.mit.edu \ elizalde@ieec.fcr.es
\ \ http://www.ieec.fcr.es/recerca/cme/eli.html} \\
Instituto de Ciencias del Espacio (CSIC) \\ \& Institut d'Estudis
Espacials de Catalunya (IEEC/CSIC) \\ Edifici Nexus, Gran Capit\`{a}
2-4, 08034 Barcelona, Spain\\ and \\ Departament ECM i IFAE,
Facultat de F\'{\i}sica, \\ Universitat de Barcelona, Diagonal 647,
08028 Barcelona, Spain \\

\vspace{16mm}

{\bf Abstract}

\end{center}

A simple model is introduced in which the cosmological constant is
interpreted as a true Casimir effect on a scalar field filling the
universe (e.g. $\mathbf{R} \times \mathbf{T}^p\times
\mathbf{T}^q$, $\mathbf{R} \times \mathbf{T}^p\times \mathbf{S}^q,
\ldots$). The effect is driven by compactifying boundary
conditions imposed on some of the coordinates, associated
with large and with small scales (the total number of large spatial
coordinates being always three). The very small
---but non zero--- value of the cosmological constant obtained from
recent astrophysical observations can be perfectly matched with
the results coming from the model, by just fixing  the
numbers of ---actually compactified--- ordinary and tiny dimensions
to be very common ones, and
being the compactification radius (for the last) in the range
$(1-10^3) \ l_{Pl}$, where $l_{Pl}$ is the Planck length. This
corresponds to solving, in a way, what has been termed by Weinberg
the {\it new} cosmological constant problem. Moreover, a
marginally closed universe is favored by the model, again in
coincidence with independent analysis of the observational
results.



\newpage

\section{Introduction}

The issue of the cosmological constant has got renewed thrust from
the observational evidence of an acceleration in the expansion of
our Universe, recently reported by two different groups
\cite{perl,ries}. There has been some controversy on the
reliability of the results obtained from those observations and on
its precise interpretation, by a number of different reasons
\cite{ries2,car}. Anyway, there is presently reasonable consensus
among the community of cosmologists that it certainly could happen
that there is, in fact, an acceleration, and that it has the order
of magnitude obtained in the above mentioned observations.
In support of this consensus, the recently issued analysis of the data
taken by the BOOMERANG \cite{boom} and MAXIMA-1 \cite{max1}
balloons have been correspondigly
crossed with those from the just mentioned observations, to
conclude that the results of BOOMERANG and MAXIMA-1 can perfectly
account for an accelerating universe and that, taking together both kinds of
observations, one inferes that we most probably live in a flat universe.
As a consequence, many theoretists have urged to try to explain this
fact, and also to try to reproduce the precise value of the
cosmological constant coming from these observations,
in the available models \cite{sts1,shs1,mon1}.

Now, as crudely stated by Weinberg in a recent review \cite{wei1},
it is even more difficult to explain why the cosmological constant
is so small but non-zero, than to build theoretical models where
it exactly vanishes \cite{wei2}. Rigorous calculations performed
in quantum field theory on the vacuum energy density, $\rho_V$,
corresponding to quantum fluctuations of the fields we observe in
nature, lead to values that are over 120 orders of magnitude in excess
of the values allowed by observations of the space-time around us.

Rather than trying to understand the fine-tuned cancellation of
such enormous values at this {\it local} level (a very difficult
question that we are going to leave unanswered, and even
unattended, here), in this paper we will elaborate on a quite
simple and primitive idea (but, for the same reason, of far
reaching, inescapable consequences), related with the {\it global} topology
of the universe \cite{ct1} and in connection with the possibility that a very
faint, massless scalar field pervading the universe could exist.
Fields of this kind are ubiquitous in inflationary models,
quintessence theories, and the like. In other words, we do not
pretend here to solve the old problem of the cosmological
constant, not even to contribute significantly to its
understanding, but just to present an extraordinarily simple model
which shows that the right order of magnitude of (some
contributions to) $\rho_V$, in the precise range deduced from the
astrophysical observations \cite{perl,ries}, e.g. $\rho_V \sim
10^{-10}$ erg/cm$^3$, are not difficult to obtain. To say it in
different words, we only address here what has been termed by
Weinberg \cite{wei1} the {\it new} cosmological constant problem.

In short, we shall assume the existence of a scalar field
background extending throught the universe and shall calculate
the contribution to the cosmological constant coming from the
Casimir energy density \cite{Casimir} corresponding to this field for some
typical boundary conditions. The ultraviolet contributions
will be safely set to zero by some mechanism of a fundamental theory.
 Another hypothesis will be the
existence of both large and small dimensions (the total number of large
spatial
coordinates will be always three), some of which (from
each class) may be compactified, so that the global topology of
the universe will play an important role, too.
There is by now a quite extense literature both in the subject of what is the
global topology of spatial sections of the universe \cite{ct1} and also on the
issue of the possible contribution of the Casimir effect as a source of
some sort of cosmic energy, as in the case of the creation of a neutron star
\cite{sokol1}. There are arguments that favor different topologies,
as a compact hiperbolic manifold for the spatial section, what would have
clear observational consequences \cite{css-mfo}.
Other interesting
work along these lines was reported in \cite{eejmp12}
and related ideas have been discussed very recently in
\cite{banks1}. However, our paper differs from all those in several respects.
To begin, the emphasis is put now in obtaining the right order of magnitude
for the effect, e.g., one that matches the recent observational results.
At the present stage, in view of the observational precission,
 it has no sense to consider the whole amount of
possibilities concerning the nature of the field, the different models
for the topology of
the universe, and the different boundary conditions possible.

At this level, from
our previous experience in these calculations and from the many tables
(see, e.g., \cite{zb1,zb2,zbas1} where precise values of the
Casimir effect corresponding to a number of different configurations have
been reported), we realize  that the range of {\it orders of
magnitude} of the vacuum energy density for the most common
possibilities is not so widespread, and may only differ by at most
a couple of digits. This will allow us, both for the sake
of simplicity {\it and} universality, to deal with a most simple
situation, which is the one corresponding to a scalar field with
periodic boundary conditions. Actually, as explained in
\cite{eenc} in detail, all other cases for parallel plates, with
any of the usual boundary conditions, can be
reduced to this one, from a mathematical viewpoint.

\section{Two basic space-time models}

Let us thus consider a universe with a space-time of one of the
following types: $\mathbf{R^{d+1}} \times \mathbf{T}^p\times
\mathbf{T}^q$, $\mathbf{R^{d+1}} \times \mathbf{T}^p\times \mathbf{S}^q,
\ldots$, which are actually plausible models for the space-time
topology. A (nowadays) free scalar field pervading the universe
will satisfy \begin{eqnarray} (-\Box +M^2) \phi =0, \end{eqnarray}
restricted by the
appropriate boundary conditions (e.g., periodic, in the first case
considered).  Here, $d\geq 0$ stands for a possible number of
non-compactified dimensions.

Recall now that the physical contribution to the vacuum or
zero-point energy $< 0 | H |  0 >$ (where $H$ is the Hamiltonian
corresponding to our massive scalar field and $|  0 >$ the vacuum
state) is obtained on subtracting to these expression ---with the
vacuum corresponding to our compactified spatial section with the
assumed boundary conditions--- the vacuum energy corresponding to
the same situation with the only change that the compactification
is absent (in practice this is done by conveniently sending the
compactification radii to infinity). As well known, both of these
vacuum energies are in fact infinite, but it is its {\it
difference}
\begin{eqnarray}
E_C = \left.  < 0 | H |  0 >\right|_R  - \left. < 0 | H |
0 >\right|_{R\rightarrow \infty}
\end{eqnarray}
(where $R$ stands here  for a typical compactification length)
that makes physical sense, giving rise to the finite value of the
Casimir energy $E_C$, which will depend on $R$ (after a well
defined regularization/renormalization procedure is carried out).
In fact we will discuss the Casimir (or vacuum) energy {\it
density}, $\rho_C=E_C/V$, which can account for either a finite or
an infinite volume of the spatial section of the universe (from
now on we shall assume that all diagonalizations already
correspond to energy densities, and the volume factors will be
replaced at the end). In terms of the spectrum $\{\lambda_n\}$ of
$H$:
\begin{eqnarray}
< 0 | H |  0 > = \frac{1}{2} \sum_n \lambda_n,
\end{eqnarray}
where the sum over $n$ is a sum over the whole spectrum, which
involves, in general, several continuum and several discrete
indices. The last appear tipically when compactifying the space
coordinates (much in the same way as time compactification gives
rise to finite-temperature field theory), as in the cases we are
going to consider. Thus, the cases treated will involve
integration over $d$ continuous dimensions and multiple summations
over $p+q$ indices (for a pedagogical description of this
procedure, see \cite{eenc}).

To be precise, the physical vacuum energy density corresponding to our case,
where the contribution of a scalar field, $\phi$ in a (partly) compactified
spatial section of the universe is considered, will be
denoted by $\rho_\phi$ (note that this is just the contribution to
$\rho_V$ coming from this field, there might be other, in general).
It is given by \begin{eqnarray}
 \rho_\phi =\frac{1}{2}
\sum_{\mbox{\bf k}} \frac{1}{\mu} \left(k^2 +M^2\right)^{1/2},
\label{c2} \end{eqnarray} where the sum $\sum_{\mbox{\bf k}}$ is a
generalized one (as explained above) and $\mu$ is the usual
mass-dimensional parameter to render the eigenvalues adimensional
(we take $\hbar =c =1$ and shall insert the dimensionfull units
only at the end of the calculation). The mass $M$ of the field
will be here considered to be arbitrarily small and will be kept
different from zero, for the moment, for computational reasons
---as well as for physical ones, since a very tiny mass for the
field can never be excluded. Some comments about the choice of our
model are in order. The first seems obvious: the coupling of the
scalar field to gravity should be considered. This has been done
in all detail in, e.g.,
 \cite{pr1} (see also the references therein). The conclusion is that
taking it into account does not change the results to be obtained
here. Of course, the renormalization of the model is rendered much
more involved, and one must enter a discussion on the orders of
magnitude of the different contributions, which yields, in the
end, an ordinary perturbative expansion, the coupling constant
being finally re-absorbed into the mass of the scalar field. In
conclusion, we would not gain anything new by taking into account
the coupling of the scalar field to gravity. Owing, essentially,
to the smallness of the resulting mass for the scalar field, one
can prove that, quantitatively, the difference in the final result
is at most of a few percent.

Another important consideration is the fact that our model is
stationary, while the universe is expanding. Again, careful
calculations show that this effect can actually be dismissed at
the level of our order of magnitude calculation, since its value
cannot surpass the one that we will get (as is seen from the
present value of the expansion rate $\Delta R /R \sim 10^{-10}$
per year or from direct consideration of the Hubble coefficient).
 As  before, for the sake of simplicity, and in order to focus just on
 the essential issues of our argument, we will perform
a (momentaneously) static calculation. As a consequence, the value
of the Casimir energy density, and of the cosmological constant,
to be obtained will correspond to the present epoch, and are bound
to change with time.

The last comment at this point would be that (as shown by the many
references mentioned above), the idea presented here is not
entirely new. However, the simplicity and the generality of its
implementation below are indeed brand new. The issue at work here
is absolutely independent of {\it any} specific model, the only
assumptions having been clearly specified before (e.g., existence
of a very light scalar field and of some reasonably compactified
scales, see later). Secondly, it will turn out, in the end, that
the only `free parameter' to play with (the number of compactified
dimensions) will actually not be that `free' but, on the contray,
very much constrained to have an admissible value. This will
become clear after the calculations below. Thirdly, although the
calculation may seem easy to do, in fact it is not so. Recently
derived reflection identities will allow us to to perform it {\it
analitically}, for the first time.

\section{The vacuum energy density and its regularization}

To exhibit explicitly a couple of the wide family of cases
considered, let us write down in detail  the formulas
corresponding to the two first topologies, as described above. For
a ($p,q$)-toroidal universe, with $p$ the number of `large' and
$q$ of `small' dimensions: \begin{eqnarray} && \hspace*{-12mm}
\rho_\phi =\frac{\pi^{-d/2}}{2^d\Gamma (d/2) \prod_{j=1}^p a_j
\prod_{h=1}^q b_h} \int_0^\infty dk \, k^{d-1} \sum_{\mbox{\bf
n}_p=-\mathbf{\infty}}^{\mathbf{\infty}} \sum_{\mbox{\bf
m}_q=-\mathbf{\infty}}^{\mathbf{\infty}} \left[ \sum_{j=1}^p
\left( \frac{2\pi n_j}{a_j}\right)^2 + \sum_{h=1}^q\left(
\frac{2\pi m_h}{b_h}\right)^2 +M^2 \right]^{1/2} \label{t1} \\ &&
\sim  \frac{1}{a^pb^q} \sum_{\mbox{\bf n}_p, \mbox{\bf
m}_q=-\mathbf{ \infty}}^{\mathbf{\infty}} \left(
\frac{1}{a^2}\sum_{j=1}^p n_j^2 + \frac{1}{b^2} \sum_{h=1}^q m_h^2
+M^2 \right)^{(d+1)/2+1},\label{t2}
\end{eqnarray} where the last formula corresponds to the case when
all large (resp. all small) compactification scales are the same.
In this last expression the squared mass of the field should be
divided by $4\pi^2\mu^2$, but we have renamed it again $M^2$ to
simplify the ensuing formulas (as $M$ is going to be very small,
we need not keep track of this change). We also will not take care
for the moment of the mass-dim factor $\mu$ in other places $-$as
is usually done$-$ since formulas would get unnecesarily
complicated and there is no problem in recovering it at the end of
the calculation. For a ($p$-toroidal, $q$-spherical)-universe, the
expression turns out to be \begin{eqnarray} \rho_\phi
&=&\frac{\pi^{-d/2}}{2^d\Gamma (d/2)\, \prod_{j=1}^p a_j \ b^q}
\int_0^\infty dk \, k^{d-1}\, \sum_{\mbox{\bf
n}_p=-\mathbf{\infty}}^{\mathbf{\infty}} \sum_{l=1}^\infty
P_{q-1}(l) \left[ \sum_{j=1}^p \left( \frac{2\pi
n_j}{a_j}\right)^2 +  \frac{Q_2(l)}{b^2} +M^2 \right]^{1/2}
\label{ts1} \\ &\sim & \frac{1}{a^pb^q} \sum_{\mbox{\bf
n}_p=-\mathbf{\infty}}^{\mathbf{\infty}} \sum_{l=1}^\infty
P_{q-1}(l)  \left( \frac{4\pi^2}{a^2}\sum_{j=1}^p n_j^2 +
 \frac{l(l+q)}{b^2}  +M^2 \right)^{(d+1)/2+1},\label{ts2}
\end{eqnarray} where $P_{q-1}(l)$ is a polynomial in $l$ of degree $q-1$,
and where the second formula corresponds to the similar situation
as the second one before. On dealing with our observable universe,
in all these expression we assume that $d=3-p$, the number of
non-compactified, `large' spatial dimensions (thus, no $d$
dependence will remain).

As is clear, all these expressions for $\rho_\phi$ need to be
regularized. We will use zeta function regularization, taking
advantage of the very powerful equalities that have been derived
recently \cite{eecmp1,ke1}, which reduce the enormous burden of
such computations to the easy application of some formulas. For
the sake of completeness, let us very briefly summarize how this
works \cite{eejpa1,eenc}. We deal here only with the case when the
spectrum of the Hamiltonian operator is known explicitly. Going
back to the most general expressions of the Casimir energy
corresponding to this case, namely Eqs. (\ref{c1})-(\ref{c2}), we
replace the exponents in them with a complex variable, $s$, thus
obtaining the zeta function associated with the operator as:
\begin{eqnarray} \zeta (s) =\frac{1}{2} \sum_{\mbox{\bf k}}
\left(\frac{k^2 +M^2}{\mu^2}\right)^{-s/2}. \label{z2}
\end{eqnarray} The next step is to perform the analytic
continuation of the zeta function from a domain of the complex
$s$-plane with Re $s$ big enough (where it is perfectly defined by
this sum) to the point $s=-1$, to obtain:
\begin{eqnarray} \rho_\phi = \zeta (-1). \end{eqnarray} The effectiveness of this method
has been sufficiently described before (see, e.g.,
\cite{zb1,zb2}). As we know from precise Casimir calculations in
those references, no further subtraction or renormalization is
needed in the cases here considered, in order to obtain the
physical value for the vacuum energy density (there is actually a
subtraction at infinity taken into account, as carefully described above,
 but it is of null value, and
no renormalization, not even a finite one, very common to other
frameworks, applies here).

Using the recent formulas \cite{eecmp1} that generalize the
well-known Chowla-Selberg expression to the situations considered
above, Eqs. (\ref{t1}) and (\ref{ts1}) ---namely,
multidimensional, massive cases--- we can provide arbitrarily
accurate results for different values of the compactification
radii. However, as argued above we can only aim here at matching
the {\it order of magnitude} of the Casimir value and, thus, we
shall just deal with the most simple case of Eq. (\ref{t2}) (or
(\ref{t1}), which yield the same orders of magnitude as the rest).
Also in accordance with this observation, we notice that among
the models here considered and which
lead to the values that will be obtained below, there are
in particular the very important typical cases of isotropic universes
with the spherical
topology. As all our discussion here is in terms of orders of magnitude and
not of precise values with small errors, all these cases are included on
{\it equal footing}. But, on the other hand, it has no sense to present a lengthy
calculation dealing in detail with all the possible spatial geometries.
Anyhow, all these
calculations are very similar to the one to be carried out here, as has been
described in detail elsewhere \cite{eejmp12,zb1,zb2}.

For the analytic continuation of the zeta function corresponding
to (\ref{t1}), we obtain \cite{eecmp1}: \begin{eqnarray}
\zeta(s)&=&\frac{2\pi^{s/2+1}}{a^{p-(s+1)/2}b^{q-(s-1)/2} \Gamma
(s/2)} \sum_{\mbox{\bf m}_q=-\mathbf{\infty}}^{\mathbf{\infty}}
\sum_{h=0}^p \left(_{\,\displaystyle h\,}^{\,\displaystyle p\,}\right) 2^h
\sum_{\mbox{\bf n}_h=1}^{\mathbf{\infty}} \left(
\frac{\sum_{j=1}^h n_j^2 }{\sum_{k=1}^q
m_k^2+M^2}\right)^{(s-1)/4} \nonumber \\ && \times K_{(s-1)/2} \left[
\frac{2\pi a}{b} \sqrt{\sum_{j=1}^h n_j^2 \left(\sum_{k=1}^q
m_k^2+M^2\right)}\right], \label{z11} \end{eqnarray} where $K_\nu (z)$ is
the modified Bessel function of the second kind. Having performed
already the analytic continuation, this expression is ready for
the substitution $s=-1$, and yields
 \begin{eqnarray}
\rho_\phi = -\frac{1}{a^pb^{q+1}} \sum_{h=0}^p \left(_{\,\displaystyle
h\,}^{\,\displaystyle p\,}\right) 2^h \sum_{\mbox{\bf
n}_h=1}^{\mathbf{\infty}} \sum_{\mbox{\bf
m}_q=-\mathbf{\infty}}^{\mathbf{\infty}} \sqrt{\frac{\sum_{k=1}^q
m_k^2+M^2}{\sum_{j=1}^h n_j^2 }} \ K_1 \left[ \frac{2\pi a}{b}
\sqrt{\sum_{j=1}^h n_j^2 \left(\sum_{k=1}^q
m_k^2+M^2\right)}\right]. \label{c11} \end{eqnarray} Now, from the behaviour
of the function $K_\nu (z)$ for small values of its argument, \begin{eqnarray}
K_\nu (z) \sim \frac{1}{2} \Gamma (\nu) (z/2)^{-\nu}, \qquad z \to
0, \label{kna} \end{eqnarray} we obtain, in the case when $M$ is very small,
\begin{eqnarray} \rho_\phi &=& -\frac{1}{a^pb^{q+1}} \left\{ M\, K_1 \left(
\frac{2\pi a}{b} M \right)+ \sum_{h=0}^p \left(_{ \, \displaystyle h\,}^{\,
\displaystyle p\,}\right) 2^h \sum_{\mbox{\bf n}_h=1}^{\mathbf{\infty}}
\frac{M}{\sqrt{\sum_{j=1}^h n_j^2 }} \ K_1\left( \frac{2\pi a}{b}
M \sqrt{\sum_{j=1}^h n_j^2} \right)\right.\nonumber \\ && + \left. {\cal
O} \left[ q\sqrt{1+M^2} K_1\left( \frac{2\pi
a}{b}\sqrt{1+M^2}\right) \right]\right\}. \label{ff0}\end{eqnarray}
At this stage, the only presence of the mass-dim parameter $\mu$
is as $M/\mu$ everywhere. This does not conceptually affect the
small-$M$ limit, $M/\mu << b/a$. Using
(\ref{kna}) and inserting now in the expression the $\hbar$ and
$c$ factors, we finally get \begin{eqnarray} \rho_\phi = -\frac{\hbar c}{2\pi
a^{p+1}b^q} \left[1+\sum_{h=0}^p \left(_{\,\displaystyle h\,}^{\,\displaystyle
p\,}\right) 2^h \alpha \right]+ {\cal O} \left[ q K_1\left(
\frac{2\pi a}{b}\right) \right], \label{ff1} \end{eqnarray} where $\alpha$
is some finite constant, computable and under control,
which is obtained as an explicit
geometrical sum in the limit $M\rightarrow 0$.
It is remarkable that we here obtain such a well defined limit,
independent of $M^2$, provided that $M^2$ is small enough. In other words,
a physically very nice situation turns out to correspond, precisely,
to the mathematically rigorous case. This is moreover, let me
repeat, the kind of expression that one gets not just for the model
considered, but for {\it many} general cases, corresponding to
different fields, topologies, and boundary conditions ---aside
from the sign in front of the formula, that may change with the
number of compactified dimensions and the nature of the boundary
conditions (in particular, for Dirichlet boundary conditions one
obtains a value in the same order of magnitude but of opposite
sign).

\section{Numerical results}

For the most common variants, the constant $\alpha$ in (\ref{ff1})
has been calculated to be of order $10^2$, and the whole factor, in
brackets, of the first term in (\ref{ff1}) has a value of order $10^7$.
This shows the value of a precise calculation, as the one undertaken here,
together with the fact that just a naive consideration of the dependences
of $\rho_\phi$ on the powers of the compactification radii, $a$ and $b$,
is {\it not enough} in order to obtain the correct result. Notice,
moreover, the non-trivial change in the power dependencies from going
from Eq. (\ref{ff0}) to Eq. (\ref{ff1}).

For the compactification radii at small scales, $b$, we
shall simply take the magnitude of the Planck length, $b \sim
l_{P(lanck)}$, while the typical value for the large scales, $a$,
will be taken to be the present size of the observable universe, $a\sim R_U$.
With this choice, the order of the quocient $a/b$ in the argument
of $K_1$ is as big as: $a/b \sim 10^{60}$. Thus, we see
immediately that, in fact, the final expression for the vacuum
energy density is completely independent of the mass $M$ of the
field, provided this is very small (eventually zero). In fact,
since the last term in Eq. (\ref{ff1}) is exponentially vanishing,
for large arguments of the Bessel function $K_1$, this
contribution is zero, for all practical purposes, what is already
a very nice result. Taken in ordinary units (and after tracing back all
the transformations suffered by the mass term $M$) the actual bound on
the mass of the scalar field is   $M \leq 1.2 \times 10^{-32}$ eV, that
is, physically zero, since it is lower by several orders of magnitude
than any bound comming from the more usual SUSY theories $-$where in fact
scalar fields with low masses of the order of that of the lightest neutrino
do show up \cite{shs1}, which may have observable implications.

\begin{table}[bht]

\begin{center}

\begin{tabular}{|c||c|c|c|c|}
\hline \hline $\rho_\phi$ & $p=0$ & $p=1$ & $p=2$ & $p=3$ \\
 \hline \hline
$b=l_P$ & $10^{-13}$ & $10^{-6}$ & 1 & $10^5$ \\ \hline $b=10\,
l_P$  &$10^{-14}$ & ($10^{-8}$) & $10^{-3}$ & $10$
 \\ \hline
$b=10^2 l_P$  &$10^{-15}$ & [$10^{-10}$] & $10^{-6}$ & $10^{-3}$
 \\ \hline
$b=10^3 l_P$ & $10^{-16}$ & ($10^{-12}$) & [$10^{-9}$] &
$10^{-7}$ \\ \hline $b=10^4 l_P$ & $10^{-17}$ & $10^{-14}$ &
($10^{-12}$) & [$10^{-11}$] \\ \hline $b=10^5 l_P$ & $10^{-18}$ &
$10^{-16}$ & $10^{-15}$  & $10^{-15}$
 \\ \hline \hline \end{tabular}

\caption{{\protect\small Orders of magnitude of the vacuum energy
density contribution, $\rho_\phi$, of a massless scalar field to
the cosmological constant, $\rho_V$, for $p$ large compactified
dimensions and $q=p+1$ small compactified dimensions,
$p=0,\ldots,3$, for different values of the small compactification
length, $b$, proportional to the Planck length $l_P$. In brackets
are the values that exactly match the observational value of the
cosmological constant, and in parenthesis the otherwise closest
approximations to that value.}}

\end{center}

\end{table}

By replacing all these values in Eq. (\ref{ff1}), we obtain the
results listed in  Table 1, for the orders of magnitude of the
vacuum energy density corresponding to a sample of different
numbers of compactified (large and small) dimensions and for
different values of the small compactification length in terms of
the Planck length. Notice again that the total number of large space
dimensions is three, as corresponds to our observable universe.
 As we see from the table, good coincidence
with the observational value for the cosmological constant is
obtained for the contribution of a massless scalar field,
$\rho_\phi$, for $p$ large compactified dimensions and $q=p+1$
small compactified dimensions, $p=0,\ldots,3$, and this for values
of the small compactification length, $b$, of the order of 100 to
1000 times the Planck length $l_P$ (what is actually a very
reasonable conclusion, according also to other approaches). To be
noticed is the fact that full agreement is obtained only for cases
where there is exactly one small compactified dimension in excess
of the number of large compactified dimensions. We must point out
that the $p$ large and $q$ small dimensions are not all that
are supposed to exist (in that case $p$ should be at least, and at most,
3 and the other cases would lack any physical meaning). In fact, as we
have pointed out before, $p$ and $q$ refer to the compactified dimensions
only, but there may be other, non-compactifed dimensions (exactly $3-p$ in
the case of the `large' ones), what translates into a slight modification
of the formulas above,  but does not change the order of magnitude of
the final numbers obtained, assuming the most common boundary conditions
for the non-compactified dimensions (see e.g. \cite{zb2} for an
explanation of this technical point). In particular, the
cases of pure spherical compactification and of mixed toroidal (for small
magnitudes) and spherical (for big ones) compactification can be treated
in this way and yield results in the same order of magnitude range. Both
these cases correspond to (observational) isotropic spatial geometries.
Also to be remarked again is the non-triviality of these calculations,
when carried out exactly, as done here, to the last expression, what
is apparent from the use of the generalized Chowla-Selberg formula.
Simple power counting is absolutely unable to provide the correct
order of magnitude of the results.

\section{Conclusions}

Dimensionally speaking, within the global approach
adopted in the present paper
everything is dictated, in the end, by the two basic lengths in
the problem, which are its Planck value and the radius of the observable
Universe. Just by playing with these numbers in the context of
our (very precise) calculation of the
Casimir effect, we have shown that the observed value of
$\rho_V$ may be remarkably well fitted, under general
hypothesis, for the most common models of the space-time topology.
Notice also that the most precise fits with the observational
value of the cosmological constant are obtained for $b$ between
$b=100 \, l_P$ and $b=1000 \, l_P$, with (1,2) and (2,3)
compactified dimensions, respectively.
The fact that the value obtained for
the cosmological constant is so sensitive to the input may be viewed as
a drawback but also, on the contrary, as
a very {\it positive} feature of our model. For one, the table has a sharp
discriminating power. In other words, there is
in fact no tuning of a `free parameter' in our model and the number of large
compactified dimensions could have been fixed beforehand, to respect what
we know already of our observable universe.

 Also, it proves that the observational
value is not easy at all to obtain. The table itself proves that there is only
very little chance of getting the right figure (a truly narrow window, since
very easily we are off by
several orders of magnitude). In fact, if we trust this value
with the statistics at hand, we can undoubtedly claim $-$through use
of our model$-$
that the ones so clearly picked up by Table 1 are {\it the} only
two  possible configurations of our observable universe
(together with a couple more coming from corresponding spherical
compactifications).
And all them
correspond to a marginally closed universe, in full agreement too
with other completely independent analysis of the observational
data \cite{car,perl,ries}.

Many questions may be posed to the simple models presented here, as concerning
the dynamics of the scalar field, its couplings with gravity and other fields,
a possible non-symmetrical behaviour with respect to the large and small
dimensions, or the relevance of vacuum polarization
(see \cite{sts2}, concerning this last point). Above we have already argued
that they can be proven to have little influence on the final numerical
result (cf., in particular, the mass obtained for the scalar field in Ref.
\cite{pr1}, extremely close to our own result,
and the corresponding discussion there).
From the very existence and specific properties
of the cosmic microwave radiation (CMB) $-$which mimics somehow the
situation described (the `mass' corresponding to the CMB is also in the
sub-lightest-neutrino range)$-$ we are led to the conclusion that such a field
could be actually present, unnoticed, in our observable universe.
In fact, the existence of scalar fields of very low masses
is also demanded by other frameworks, as SUSY models, where
the scaling behaviour of the cosmological constant has been
considered \cite{shs1}.

Let us finally recall that the Casimir effect is an ubiquitous phenomena.
Its contribution may be small (as it seems to be the case, yet controverted,
to sonoluminiscence), of some 10-30$\%$ (that is, of the right order of
magnitude, as in wetting phenomena involving He in condensed matter physics),
or even the whole thing (as in recent, dedicated experimental confirmations
of the effect). Here we have seen that it provides a contribution of the
right order of magnitude, corresponding to our present epoch in the evolution
of the universe. The implication that this calculation bears for the early
universe and inflation is not clear from the final result, since it should be
adapted to the situation and boundary conditions corresponding to those
primeval epochs, what cannot be done straightforwardly.
Work along this line is in progress.

\medskip

\noindent{\bf Acknowledgments}

I am grateful to Robert Kirshner, Tom Mongan, Varun Sahni and Joan Sol\`a for
important comments. Thanks are also given to the referee for interesting
suggestions that have led to an improvement of the paper.
This investigation has been supported by DGICYT (Spain), project
PB96-0925 and by CIRIT (Generalitat de Catalunya),  grants
1997SGR-00147 and 1999SGR-00257.

\vspace{1mm}


\end{document}